\documentstyle[12pt,epsfig,amstex]{article}
\chardef\anciennecat=\catcode`\@
\catcode`\@=11
\renewcommand\section{\@startsection {section}{1}{\z@}%
                                   {-3.5ex \@plus -1ex \@minus -.2ex}%
                                   {2.3ex \@plus.2ex}%
                                   {\reset@font\large\bfseries}}
\catcode`\@=\anciennecat
\begin{document}

\title{High energy factorization predictions
for the charm structure function $F_2^c$ at HERA}
\author{S. Munier and R. Peschanski\\
CEA, Service de Physique Th\'eorique,
CE-Saclay\\
F-91191 Gif-sur-Yvette Cedex, France}
\maketitle

\begin{abstract}
High energy factorization predictions for $F_2^c$ are derived
using BFKL descriptions of the proton structure function $F_2$ at HERA.
The model parameters are fixed by
a fit of $F_2$ at small $x$.
Two different approaches of
the non perturbative proton input are shown to
correspond to the factorization at the gluon or quark level, respectively.
The predictions for $F_2^c$ are in agreement with the data within the present
error bars. However,
the photon wave-function formulation (factorization at quark level)
predicts significantly higher $F_2^c$ than both gluon factorization and a next-leading
order DGLAP model.
\end{abstract}

\clearpage

\section{Motivation} 

High energy factorization \cite{catani,ellis} is a QCD
factorization scheme suited for high-energy hard processes - and in particular
for deep-inelastic $e^\mp\!\!-\!\!p$ scattering 
in the small
$x$-regime ($x\simeq \frac{Q^2}{W^2}$)
($W^2$ is the energy squared in the  virtual photon-proton center of mass frame). 
This scheme takes into account the
resummation of the $\left(\alpha_s \log \frac{1}{x}\right)^n$
terms in the QCD perturbative expansion of the structure functions.
It amounts to proving the factorization of the perturbative amplitude in terms of a
gluon-gluon BFKL kernel \cite{bfkl} convoluted 
with a first order (virtual)photon-(virtual)gluon cross section.
A graphical illustration of this property is given in figure 1.
The main difference between this scheme and the renormalization group factorization 
\cite{collins} which is valid at finite $x$  is that the former 
involves a bidimensional integration
in transverse momentum $k_\bot$, whereas the latter is a convolution in energy.
Interestingly enough, high energy factorization can be applied to the resummation of
the $\left(\alpha_s \log \frac{W^2}{M^2}\right)^n$ 
terms in heavy quark pair photoproduction, and 
the combined resummation for heavy quark leptoproduction involving
$\left(\alpha_s\log\frac{W^2}{Q^2}\right)^n$ 
and $\left(\alpha_s\log\frac{W^2}{M^2}\right)^n$ 
terms can also be studied \cite{catani}. Our aim is to test 
the high energy factorization
predictions for the massive charm quark pair contribution to
$F_2$ at HERA.

In references \cite{npr,nprw},
a formulation and a phenomenological analysis of the proton structure function $F_2$ 
satisfying high energy factorization in the framework of Mueller's
color dipole picture \cite{mueller} of BFKL dynamics were proposed.
$F_2$ is the triple transverse momentum convolution of a coefficient function,
the BFKL kernel and a non perturbative term.
The non perturbative inputs, which can be interpreted as the density and size of
primary dipoles in the parton sea of the proton \cite{nprw}, 
lead to a satisfactory description
of $F_2$ at HERA range, with 3 parameters.
In this analysis, quarks were assumed massless, and the convolution integral
was approximated by a steepest-descent method for sake of simplicity.
However, since the coefficient functions
were considered constant and the quarks massless, 
the procedure was not specifically  a check of 
high energy factorization itself.

In the following, we shall discuss the high energy factorization predictions
by taking full account of the convolution integral
and of the massive quark component in the phenomenological analysis.

On the experimental point of view, 
the HERA experiments have recently published \cite{f2c}
data for the contribution of charmed meson production 
to the structure function $F_2$. Their analysis was based on 
$D^0$ and $D^*$ meson tagging.
This allows to single out the charm contribution $F_2^c$ to the total structure
function and thus to investigate the quark mass dependence of the
structure functions.
We use this nice possibility to check whether the high energy factorization
scheme gives a correct prediction for the mass-dependent contribution
both in $x$ and $Q^2$.
Our analysis can be considered as the analog within the BFKL dynamics of
the one which can be performed using ordinary renormalization group evolution
\cite{GRV,thorne}. Thus, we are also aiming to see whether there are 
agreements or
differences in charm leptoproduction between the
different schemes available for analyzing deep-inelastic scattering.

However, the high energy factorization scheme adapted to the proton structure functions
is not uniquely defined in terms of the separation between perturbative and
non-perturbative contributions. Indeed, an {\it a priori} different factorization
exists in which the virtual photon is described in terms of
a quark-antiquark wave function configuration which then interacts with the proton (see for
instance \cite{bjorken,nikolaev}).
One of our goals is to investigate the possible differences between the
two schemes.
Recently also, some theoretical doubt has been cast on the validity of
the operator product expansion in the low-$x$ range due to the $k_\bot$-diffusion
property \cite{bartes} inherent to the BFKL dynamics.
In the context of our structure function studies, we thus also want
to discuss phenomenologically the kinematical range
in $x$ and $Q^2$ where high energy factorization is valid and not spoiled by 
$k_\bot$-diffusion in the low momentum region.

The plan of our study is the following.
Section {\bf 2} recalls the formulation of high energy factorization
for the pair production of quarks of mass $M$ in terms of
the coefficient functions and unintegrated gluon distribution in
the proton.
We give the $M$-dependent expression of the coefficient functions.
In section {\bf 3}, we derive the expression of the proton structure function
including the charm component in the framework of high energy factorization.
We derive a new constraint on the unintegrated gluon distribution
function due to the renormalization group property at high $k_\bot$.
In section {\bf 4}, we consider the alternative model of factorization
based on the wave function formulation. We show the connection between the 
two models. We prove the equivalence
at perturbative level between both schemes for the hard vertex while they
are based on different non-perturbative inputs.
In section {\bf 5}, 
both models are applied to a phenomenological fit of $F_2$ and parameter
free predictions for $F_2^c$, $F_L$, and $F_L^c$. 
The final section {\bf 6} contains our conclusions on the validity
of the high energy factorization predictions at HERA and an outlook on
future studies.
Taking into account both present experimental and theoretical uncertainties,
high energy factorization is proved to be in agreement with the present experimental
data. However, more precise data could discriminate between
the different schemes, since the photon wave function formulation predicts higher
$F_2^c$ than both the gluon factorization and a next-leading order DGLAP model.


\section{High energy factorization}  

Let us compute the pair production of quarks
of arbitrary mass $M$ at the
virtual photon vertex of deep-inelastic scattering (see figure 1).
Using the high energy factorization scheme \cite{catani}, the inclusive
transverse (resp. longitudinal) structure functions 
$F_T$ ($F_L$) can be expressed as follows:
\begin{multline}
F_{T,L}(Y,Q^2,M^2) =
\frac{1}{4\pi^2\alpha_{em}}\frac{Q^2}{4{M}^2}
\int d^2k_\bot\int_0^\infty dy\times\\
   \hat{\sigma}_{\gamma^* g,T,L}
(Y\!\!-\!\!y,q_\bot/M,k_\bot/M)\;{\mathcal F}(y,k_\bot)\ ,
\label{eqn:0}
\end{multline}
where $Q^2=-q^2$ is the virtuality of the photon.
$Y$ represents the rapidity range available for the reaction. 
${\mathcal F}(y,k_\bot)$ is the unintegrated gluon distribution
\cite{catani}, which describes the probability of finding a gluon
with a longitudinal momentum fraction $z=e^{-y}$ and 
two-dimensional transverse momentum $k_\bot$ in the target. 
$q_\bot$ is the photon transverse momentum in the photon-proton centre
of mass frame.
$\hat{\sigma}_{\gamma^* g,T,L}$ is the hard cross section for 
(virtual)photon-(virtual)gluon fusion computed at order $\alpha_s\alpha_{em}$.

In order to express more easily high energy factorization, 
it is convenient to work with triple Mellin-transforms with respect
to the rapidity $Y$ and transverse momenta $k_\bot$ and $q_\bot$.
The result is:
\begin{multline}
F_{T,L}(Y,Q^2,M^2)=
\frac{1}{4\pi^2\alpha_{em}}\frac{Q^2}{4M^2}
\int\frac{d\gamma^\prime}{2i\pi\gamma^\prime}
\left(\frac{4M^2}{Q^2}\right)^{\gamma^\prime}
\int\frac{d\gamma}{2i\pi\gamma}
    \left(\frac{4M^2}{Q_f^2}\right)^{\gamma}\\
  \int\frac{dN}{2i\pi} e^{YN}\;
h_{T,L,N}(\gamma^\prime,\gamma)\; {\mathcal F}_N(\gamma;Q_f^2),
\label{eqn:1}
\end{multline}
where $Q_f^2$ stands for the factorization scale and
the inverse Mellin-transforms in $\gamma$, $\gamma^\prime$ and $N$
are given by complex integrals along the axes
$\frac{1}{2}\pm i\infty$, $\frac{1}{2}\pm i\infty$ and $N_0\pm i\infty$ ($N_0$
larger than the real part of the rightmost singularity in the $N$-plane) respectively.
By definition, 
$h_{T,L,N}(\gamma^\prime,\gamma)$
is the triple inverse Mellin-transform of the hard $\gamma^*$-gluon cross-section,
namely:
\begin{multline}
h_{T,L,N}(\gamma^\prime,\gamma)\equiv
\gamma^\prime\gamma\int\frac{d^2q_\bot}{\pi q_\bot^2}\left(
\frac{q_\bot^2}{4M^2}\right)^{\gamma^\prime}\int
\frac{d^2k_\bot}{\pi k_\bot^2}\left(
\frac{k_\bot^2}{4M^2}\right)^\gamma\times\\
 \int_0^\infty dY e^{-YN}\;\hat{\sigma}_{\gamma^* g,T,L}(Y,q_\bot/M,k_\bot/M)\ .
\label{eqn:defhtl}
\end{multline}

\noindent
${\mathcal F}_N(\gamma;Q_f^2)$ is obtained from the 
unintegrated gluon distribution ${\mathcal F}(y,k_\bot)$.
It reads:
\begin{align}
{\mathcal F}_N(\gamma;Q_f^2) &\equiv
 \int_0^\infty dy e^{-yN}
\int\frac{d^2k_\bot}{\pi k_\bot^2}\left(
\frac{k_\bot^2}{Q_f^2}\right)^{-\gamma} {\mathcal F}(y,k_\bot)\ .
\label{eqn:unint}
\end{align}
Note that
the coefficient functions $h_{T,L,N}$ are known to
have a weak dependence on $N$ \cite{catani}, hence we will
consider in the following only their values at $N=0$ denoted $h_{T,L}(\gamma;M^2)$.

The final expression for the high energy factorized structure function is
\begin{equation}
F_{T,L}(Y,Q^2,M^2)=
\int\frac{d\gamma}{2i\pi}\left(\frac{Q^2}{Q_f^2}
\right)^\gamma h_{T,L}(\gamma;M^2)
\ \frac{{\mathcal F}(Y,\gamma;Q_f^2)}{\gamma}\ ,
\label{eqn:finale}
\end{equation}
where
\begin{equation}
{\mathcal F}(Y,\gamma;Q_f^2)=\int\frac{dN}{2i\pi}e^{YN}{\mathcal F}_N(\gamma;Q_f^2)
\label{eqn:sf}
\end{equation}
and the coefficient functions related to $F_T$ and $F_L$ respectively
are \cite{catanew}:
\begin{multline}
h_T(\gamma;M^2)=\frac{\alpha_s}{6\pi} 
\frac{4^{-\gamma}}{(1+2\gamma)(1-\frac{2}{3}\gamma)}\frac{\Gamma(1\!+\!\gamma)
\Gamma^3(1\!-\!\gamma)}{\Gamma(2-2\gamma)}\frac{1}{1+\frac{4M^2}{Q^2}}
\times\\
 \left\{\left(\frac{4M^2}{Q^2}\right)^\gamma
\left((3\gamma-1)+\frac{4M^2}{Q^2}(\gamma-2)\right)\right.
+\left(1\!+\!\frac{4M^2}{Q^2}\right)^{\gamma-1}\!\!\times\\
 \left(2(1\!+\!\gamma)(2\!-\!\gamma)\!+\!
\frac{4M^2}{Q^2}(7\!+\!\gamma\!-\!6\gamma^2\!-\!
\frac{4M^2}{Q^2}(\gamma\!-\!2))\right)\times\\
  \left.{\sideset{_2}{_1}F}\left(1\!-\!\gamma,\frac{1}{2};\frac{3}{2};\frac{1}
{1+\frac{4M^2}{Q^2}}\right)\right\}
\label{eqn:coeft}
\end{multline}
and
\begin{multline}
h_L(\gamma;M^2)
 =\frac{\alpha_s}{6\pi}\frac{4^{-\gamma}}{(1+2\gamma)
(1-\frac{2}{3}\gamma)}\frac{\Gamma(1\!+\!\gamma)\Gamma^3(1\!-\!\gamma)}
{\Gamma(2-2\gamma)}\frac{1}{1+\frac{4M^2}{Q^2}}\times\\
  \left\{\left(\frac{4M^2}{Q^2}\right)^\gamma
(2(1\!-\!\gamma)+3\frac{4M^2}{Q^2})
+\left(1+\frac{4M^2}{Q^2}\right)^{\gamma-1}\right.\times\\
  \left.\left(4\gamma(1\!-\!\gamma)
-\frac{4M^2}{Q^2}(3\frac{4M^2}{Q^2}+4(1\!-\!\gamma))\right)\right.\times\\
 \left.{\sideset{_2}{_1}F}\left(1\!-\!\gamma,\frac{1}{2};\frac{3}{2};
\frac{1}{1+\frac{4M^2}{Q^2}}\right)\right\}\ ,
\label{eqn:coefl}
\end{multline}
where ${\sideset{_2}{_1}F}(a,b;c;\zeta)$ is the hypergeometric function \cite{grad}.

It is easy to check that in both limits
$M^2\rightarrow 0$ (light quarks \cite{catahaut}) and $Q^2\rightarrow 0$
(heavy flavour photoproduction \cite{catani}), the well-known expressions for the 
coefficient functions are recovered.

Inserting the expressions (\ref{eqn:coeft},\ref{eqn:coefl}) 
for $h_{T,L}(\gamma;M^2)$ and a model
for ${\mathcal F}(Y,\gamma;Q_f^2)$ in formula (\ref{eqn:finale}), 
and summing over all active flavours
weighted by their electric charges squared,
we obtain the explicit expression for the proton structure functions.


\section{The proton structure functions}

Let us now introduce the QCD dipole model for the proton structure functions.
This model satisfies high energy factorization and gives a phenomenological
description of the unintegrated structure function ${\mathcal F}_N(\gamma;Q_f^2)$,
see (\ref{eqn:unint}),
in the framework of BFKL dynamics.
We parametrize the factor
\begin{equation}
\frac{{\mathcal F}_N(\gamma;Q_f^2)}{\gamma}=\frac{\omega_N(\gamma;Q_f^2)}
{N-\frac{\alpha_sN_c}{\pi}\chi(\gamma)}\ .
\label{eqn:sing}
\end{equation}
Notice the factor $1/\gamma$ 
which corresponds to the $k_\bot$ integration
in Mellin transform and thus, $\omega_N(\gamma;Q_f^2)$ appears as the
residue of the BFKL pole in the integrated gluon distribution in the target.
${\mathcal F}_N$ is assumed to contain the well-known BFKL singularity \cite{bfkl}
at $N=\frac{\alpha_sN_c}{\pi}\chi(\gamma)$
with
\begin{equation}
\chi(\gamma)=2\Psi(1)-\Psi(\gamma)-\Psi(1\!\!-\!\!\gamma)\ .
\label{eqn:kernel}
\end{equation}

\noindent
The function $\omega_N(\gamma;Q_f^2)$ will explicitly depend
on the nature of the target. It may also contain other singularities than
the BFKL pole, which
will be discussed later on. For a target of small size (parton, massive
onium), it can be deduced from perturbative calculations, whereas for
extended targets like a proton, it is supplied by a model.

Following the suggestion of ref.~\cite{nprw}, one assumes for the
proton structure functions the following scaling
form:
\begin{equation}
\omega_{N\simeq 0}(\gamma;Q_f^2)=
   \omega(\gamma)\left(\frac{Q_f}{Q_0}\right)^{2\gamma}\ ,
\label{eqn:omega}
\end{equation}
where $\omega(\gamma)$ and $Q_0$ are the non-perturbative inputs
of the model.
Indeed, the scaling assumption (\ref{eqn:omega}) is the simplest one
allowing to obtain formulae independent of the arbitrary factorization
scale $Q_f$ (see equation (\ref{eqn:finale})). Note that $Q_0$ can be considered
to be independent of the quark mass $M$, which is quite a reasonable assumption
for a non perturbative proton parameter.
Inserting (\ref{eqn:sing}) and the expressions 
for the coefficient functions 
(\ref{eqn:coeft},\ref{eqn:coefl}) in
equation (\ref{eqn:finale}), and taking into account the scaling
(\ref{eqn:omega}), the overall formula reads:
\begin{align}
F_{T,L}(x,Q^2;Q_0^2)&=\sum_{\mathcal Q} e^2_{\mathcal Q} \; F_{T,L}
      (Y_{\mathcal Q},Q^2,M_{\mathcal Q}^2;Q_0^2)\nonumber\\
     &=\int\!\frac{d\gamma}{2i\pi}\left(\frac{Q^2}{Q_0^2}
      \right)^\gamma\!\omega(\gamma)\!
      \int\frac{dN}{2i\pi}\frac{\sum_{\mathcal Q} e^2_{\mathcal Q}
             e^{Y_{\mathcal Q}N}h_{T,L}(\gamma;M^2_{\mathcal Q})}
        {N-\frac{\alpha_sN_c}{\pi}\chi(\gamma)}\ ,
\label{eqn:tot}
\end{align}
where 
\begin{equation}
Y_{\mathcal Q}=\log\frac{1}{x(1+\frac{4M_{\mathcal Q}^2}{Q^2})}
\end{equation} 
is the maximal available rapidity range for the produced gluons in
association with quarks of mass $M_{\mathcal Q}$.
The summation $\sum_{\mathcal Q}$ takes into account the 
contributions of all active flavours
with charge $e_{\mathcal Q}$, mass $M_{\mathcal Q}$.

The unintegrated gluon distribution in the proton ${\mathcal F}_N(\gamma;Q_f^2)$
is thus model dependent in particular through the input function
$\omega(\gamma)$.
However, let us show that it obeys
a theoretical constraint when $\gamma$ goes to infinity \cite{private}. Indeed,
as we shall demonstrate,
this limit corresponds to a situation in which the intermediate gluon
emitted from the dipole (see fig.1) has a large transverse
momentum $k_\bot$. Hence its evolution from $Q_0^2$ up to $k_\bot^2$ is
governed by the gluon-gluon DGLAP evolution equation (in its
small-$x$ approximation).

Let us then consider formula (\ref{eqn:unint}) and single out the integration region
$k_\bot^2\gg Q_0^2\ $. In this hard region where $k_\bot$-ordering 
of intermediate gluons is valid, the unintegrated gluon distribution
is simply related to ${\mathcal G}(Y,k_\bot^2)$, the gluon distribution function
in the proton at the scale $k_\bot^2$, by:
\begin{equation}
{\mathcal F}(Y,k_\bot^2)=\frac{1}{\pi}\frac{d}{dk_\bot^2}
 {\mathcal G}(Y,k_\bot^2)\ .
\label{eqn:12}
\end{equation}
Introducing the Mellin transform in rapidity, the gluon-gluon 
DGLAP evolution equation at small $x$ (small $N$) reads
\begin{equation}
\int_0^\infty dYe^{-YN}{\mathcal G}(Y,k_\bot^2) \equiv
  {\mathcal G}_N(k_\bot^2)
   \simeq\left(\frac{k_\bot^2}{Q_0^2}\right)^{\frac{\alpha_s N_c}{\pi N}}
   {\mathcal G}_N(Q_0^2)\ .
\label{eqn:15} 
\end{equation}
Inserting formula (\ref{eqn:15}) into equation (\ref{eqn:unint}) 
(for $k_\bot^2\gg Q_0^2$), one gets a contribution
\begin{equation}
{\mathcal F}_N(\gamma;Q_f^2) \underset{|\gamma|\rightarrow\infty}
  {\simeq}
\left(\frac{Q_f^2}{Q_0^2}\right)^{\frac{\alpha_sN_c}{\pi N}}
\frac{1}{\gamma-\frac{\alpha_sN_c}{\pi N}}\;{\mathcal G}_N(Q_0^2)\ ,
\label{eqn:contrainte}
\end{equation}
where $\gamma=\frac{\alpha_sN_c}{\pi N}$ is the well-known DGLAP
singularity when $N$ goes to $0$.

As well-known, when $\gamma$ is small $\chi(\gamma)\simeq 1/\gamma$ 
and the BFKL pole
(\ref{eqn:sing}) can be approximated by
\begin{equation}
\frac{1}{N-\frac{\alpha_sN_c}{\pi}\chi(\gamma)}\underset
  {\gamma\rightarrow 0}{\simeq}
\frac{1}{\gamma-\frac{\alpha_sN_c}{\pi N}}\frac{\gamma}{N}
\label{eqn:16}
\end{equation} 
and thus coincides with the DGLAP pole (\ref{eqn:contrainte}).
However, in the large $|\gamma|$ region,
$|\chi(\gamma)|\simeq 2\log |\gamma|$ and thus the DGLAP pole at
$\gamma=\frac{\alpha_sN_c}{\pi N}$ is separated from the BFKL
singularity and dominates
over it at small $N$.

Integrating over this dominant singularity in the inverse 
Mellin transform (\ref{eqn:sf}),
we observe that 
the unintegrated structure function ${\mathcal F}$
satisfies the constraint
\begin{align}
\left|\left(\frac{Q_0^2}{Q_f^2}\right)^\gamma
\frac{{\mathcal F}(\gamma,Y;Q_f^2)}{\gamma}\right| 
&\underset{|\gamma|\rightarrow\infty}{\simeq}
 \left|\frac{\alpha_sN_c}{\pi\gamma^3}
 e^{Y\frac{\alpha_sN_c}{\pi\gamma}}
{\mathcal G}_{\frac{\alpha_sN_c}{\pi\gamma}}(Q_0^2)
 \right|\nonumber\\
  &\sim\frac{\alpha_sN_c}{\pi|\gamma|^3}
\ {\mathcal G}_{N\simeq 0}(Q_0^2) 
\label{eqn:cont2}
\end{align}
assuming a regular input
${\mathcal G}_{N\simeq 0}(Q_0^2)$.
On the other hand, the large $|\gamma|$ behaviour of the coefficient functions
is given by
\begin{align}
h_{T,L}(\gamma;M^2) &\underset
  {\gamma\rightarrow \infty}{\simeq}  \frac{\Gamma(1\!+\!\gamma)\Gamma^3(1\!-\!\gamma)}
  {\Gamma(2-2\gamma)}\> {\sideset{_2}{_1}F}
     (1\!-\!\gamma,1/2;3/2;1)\times\nonumber\\
   &\ \ \ \ \ \ \ \ \ \ \ \ \ \ \ \times \left\{\mbox{terms of the form}\ 
	\gamma^\alpha \left(\frac{4M^2}{Q^2}\right)^\beta 
  \right\}\nonumber\\
 &\sim  e^{-\pi|\gamma|}\times\left\{\mbox{Power-like terms}\right\}\ ,
\label{eqn:behav}
\end{align}
and thus gives an exponential cutoff at $\gamma\simeq\frac{1}{\pi}$.

\noindent
The following remarks are in order:
\begin{itemize}
\item The obtained behaviour (\ref{eqn:contrainte})
for ${\mathcal F}_N(\gamma;Q_f^2)$ at $|\gamma|$ large is in
agreement with the scaling assumption (\ref{eqn:omega})
at the pole $\gamma=\frac{\alpha_sN_c}{\pi N}$. 
\item Most importantly, the large $|\gamma|$ behaviour of the coefficient
functions $\left|h_{T,L}(\gamma,M^2)\right|
 \simeq e^{-\pi|\gamma|}$
dominates over the unintegrated structure function 
${\mathcal F}(Y,\gamma;Q_f^2)$ which is only power-like
in $\gamma$, showing that the
main integration region is for finite $|\gamma|\leq\frac{1}{\pi}$.
\item The large $|\gamma|$ behaviour of ${\mathcal F}(Y,\gamma;Q_f^2)$ 
is actually not dominated by
the BFKL singularity but by the DGLAP singularity which differ in this
domain.
However, we shall neglect this modification occurring in a domain
where the integrand is cutoff by the coefficient functions.
A more detailed analysis of high energy factorization
could need taking care of this modification.
\end{itemize}


\section{High energy factorization at the quark level}

In the previous section, we have found that the main integration region
is for finite $|\gamma|$. In fact, the structure of the integrand singularity
at $\gamma=0$ appears to be essential both for the theoretical analysis and
the phenomenological application. 
We note that in
formulae (\ref{eqn:coeft}) and (\ref{eqn:coefl}) 
for the coefficient functions,
${\sideset{_2}{_1}F}(1\!-\!\gamma,\frac{1}{2};\frac{3}{2};{1}/(1+{4M^2}/{Q^2}))
\propto 1/\gamma$ when $\gamma$
goes to $0$. We thus find that the Mellin-transform $h_{T,L,N}(\gamma;M^2)/
{\gamma}$ of $\hat{\sigma}$ in formula (\ref{eqn:0})
has in general a double pole at $\gamma=0$ (but for e.g. $h_L$ when $M^2=0$).
It is easy to realize that this double pole corresponds to an extra
$(\log{k_\bot^2}/{Q^2})$ in the small $k_\bot$-behaviour of the 
hard photon-gluon cross section due to the quark propagator (see fig.1).
This behaviour is thus characteristic of the high energy factorization formalism
at the perturbative level.

However, 
it is well-known \cite{nouvelle} that there is an ambiguity
in the separation between perturbative and non perturbative contributions
in the small-$k_\bot$ domain. The relevance of the perturbative 
double pole depends on the physical picture of the
non perturbative input.
For definiteness, we will consider
two classes of models relying on different hypotheses on the behaviour of
the residue function
$\omega(\gamma)$ near $\gamma=0$. In the first type of models, 
with 
\begin{equation}
\omega(\gamma)\underset{\gamma\rightarrow 0}
  {\simeq}\mbox{(constant)}\ ,
\end{equation}
one keeps the full perturbative information on
the $\gamma^*$-gluon vertex.
This corresponds to the {\it factorization at the gluon level}
(see fig.1). 
In the second class of models, the perturbative singularity at the hard vertex is
smoothened by the proton scaling function (\ref{eqn:omega}), namely
\begin{equation}
\omega(\gamma)\underset{\gamma\rightarrow 0}
  {\simeq}\mbox{(constant)}\times\gamma\ .
\label{eqn:cst2}
\end{equation}
The resulting single pole at the hard vertex may be interpreted as a
direct pointlike coupling of the virtual photon to the quark. This
may be interpreted as a {\it factorization at the quark level} (see fig.1).

In summary, either the contribution of the off-shell gluon
at the hard vertex is maintained, or it is compensated by the non perturbative
input. In the first case (model 1), the hard photon is assumed to probe
the gluon content of the target, and consequently, the ${1}/{\gamma^2}$ singularity
of the coefficient functions is preserved. In the latter case (model 2),
the photon probes the quark distribution and thus the coefficient function
double pole can be compensated by the proton scaling function
$\omega(\gamma)\propto\gamma$ when $\gamma$ goes to $0$, see formula (\ref{eqn:tot}).

As we shall see, a prototype of model 2 is provided by the wave function
formulation of the photon-proton interaction \cite{bjorken,nikolaev}.
In this framework,  deep-inelastic 
scattering processes are formulated in terms of the probability distribution of a
${\mathcal Q}\bar{\mathcal Q}$ pair (considered as a dipole configuration) 
in the virtual photon, 
convoluted by the dipole-proton cross section. In our case, 
the dipole-proton cross-section is described by the convolution of 
the probability distribution of primordial dipoles in the proton times
the dipole-dipole BFKL cross-section \cite{mueller}.
This will allow a direct comparison between
model 1 and 2, which have a similar parametrization
differing only by the pole structure at $\gamma=0$.

In references \cite{bjorken,nikolaev}, one finds the expressions
for the wave function and probability distribution of 
the photon ${\mathcal Q}\bar{\mathcal Q}$ states.
The virtual photon can be described in terms of probability distributions
(when the interference terms \cite{bp} are not relevant) 
depending on the quark mass $M$ and charge $e$
\begin{align}
\Phi^\gamma_T(z,r;Q^2,M^2)&=\frac{\alpha_{em}N_c}{2\pi^2}
e^2\left((z^2+(1-z)^2)\epsilon^2K_1^2(\epsilon r)
+M^2 K_0^2(\epsilon r)\right)\ ,\nonumber\\
\Phi^\gamma_L(z,r;Q^2,M^2)&=\frac{\alpha_{em}N_c}{2\pi^2}
4e^2 Q^2 z^2 (1-z)^2 K_0^2(\epsilon r)\ ,
\end{align}
where $\epsilon^2=z(1-z)Q^2+M^2$, and the
$K_{0,1}$ are the Bessel functions of second kind \cite{grad}.
$\Phi^\gamma_{T,L}(z,r;Q^2,M^2)$ are the probability distributions 
of finding a dipole configuration of
transverse size $r$ at a given $z$, the variable 
$z$ (resp. $(1-z)$) being the photon light-cone momentum
fraction carried by the antiquark (resp. quark).

The transverse and longitudinal total cross sections
$\sigma_{T,L}$ read
\begin{equation}
\sigma_{T,L}=\int d^2r dz\;
\Phi^\gamma_{T,L}(r,z;Q^2,M^2)\int d^2r_p dz_p\; 
\Phi^p(r_p,z_p)\>\sigma_{d}(r,r_p;Y)\ ,
\label{eqn:ftl}
\end{equation}
where we have introduced the probability distributions $\Phi^p(r_p,z_p)$
of dipoles inside the proton \cite{npr,nprw,bialas2}.
The dipole-dipole cross section $\sigma_d(r,r_p;Y)$ is assumed not to 
depend on $z$. 
This hypothesis means that we neglect sub-asymptotic effects related to the
momentum carried by the quarks (while we do not neglect the quark masses).
In practice, it corresponds to consider $N\simeq 0$ in the Mellin-transform
with respect to rapidity as in section {\bf 2}.

$\sigma_d(r,r_p;Y)$ 
reads \cite{mueller,bialas2,unpub}:
\begin{equation}
\sigma_{d}(r,r_p;Y)=4\pi r r_p \int\frac{d\gamma}{2i\pi}
\left(\frac{r_p}{r}\right)^{2\gamma-1}
e^{\frac{\alpha_s N_c}{\pi}\chi(\gamma)Y}A_{el}(\gamma)\ ,
\end{equation}
where $\chi(\gamma)$ is the BFKL kernel (\ref{eqn:kernel}) and
the elementary two-gluon exchange amplitude is given by
\begin{equation}
A_{el}(\gamma)=\frac{\alpha_s^2}{16\gamma^2(1\!-\!\gamma)^2}\ .
\label{eqn:ael}
\end{equation}

We define the non-perturbative scale $Q_0$ characterizing the
average dipole size by \cite{bialas2}
\begin{equation}
\int d^2 r_p r_p^{2\gamma}\int dz_p\;\Phi^p(r_p,z_p)\equiv 
  \frac{n_{eff}(\gamma)}{(Q_0^2)^\gamma}\ ,
\label{eqn:22}
\end{equation}
where $n_{eff}$ can be interpreted as the $\gamma$-dependent
average number of primary dipoles in the
proton. Finally, the Mellin-transform of the photon wave-function
is defined by:
\begin{equation}
\int \frac{d^2r}{2\pi} (r^2)^{1\!-\!\gamma}\int dz
\>\Phi_{T,L}^\gamma(r,z;M^2)=\phi_{T,L}(\gamma;M^2)
  \left(Q^2\right)^{\gamma-1}\ ,
\label{eqn:defphi}
\end{equation}
where we have explicitely factorized the photon scale dependence.
After plugging these formulae into equation (\ref{eqn:ftl}), and performing
the integrations with respect to $r$, $r_p$ and $z$, $z_p$ one finds:
\begin{equation}
\sigma_{T,L}=
  \frac{32\pi^2}{Q^2}\int\frac{d\gamma}{2i\pi}
  \left(\frac{Q^2}{Q_0^2}\right)^\gamma\!\! 
  e^{\frac{\alpha_sN_c}{\pi}\chi(\gamma)Y}\phi_{T,L}(\gamma;M^2)A_{el}(\gamma)
  n_{eff}(\gamma)\ .
\end{equation}

\noindent
Inserting the virtual photon probability distribution $\Phi^\gamma_{T,L}(r,z;Q^2,M^2)$
into equation (\ref{eqn:defphi}) and after a straightforward but tedious calculation,
one finds expressions for $\Phi_{T,L}(\gamma;M^2)$ which can be cast into the following form:
\begin{equation}
\phi_{T,L}(\gamma;M^2)=\frac{\alpha_{em}e^2}{\alpha_s}\frac{N_c}{4\pi}
  \frac{h_{T,L}(\gamma;M^2)}{\gamma}\;\left\{2^{-2\gamma+3}(1\!-\!\gamma)^2
   \frac{\Gamma(1\!-\!\gamma)}{\Gamma(\gamma)}\right\}\ ,
\label{eqn:phi}
\end{equation}
where $h_{T,L}(\gamma;M^2)/\gamma$ is related to the Mellin transform
of $\hat{\sigma}$
accounted for in the preceding section (see equation (\ref{eqn:defhtl})).
It is clear from formula (\ref{eqn:phi}) 
that the double pole of $h_{T,L}(\gamma;M^2)/{\gamma}$ is turned into 
a single pole due to the factor $\{...\}$. This shows that
the factorization of the probability distributions $\phi_{T,L}(\gamma;M^2)$ at the
hard vertex leads to the singularity structure (\ref{eqn:cst2}) of model 2.

\noindent
Some comments are in order:
\begin{itemize}
\item The $\gamma$-dependent but $M^2$-independent factor $\{...\}$ is nothing but
the coupling of the virtual gluon to the ${\mathcal Q}\bar{\mathcal Q}$ pair
configuration of the virtual photon wave function. Indeed, the two-gluon exchange
elementary amplitude (\ref{eqn:ael}) can be rewritten as
\begin{equation}
A_{el}(\gamma)=\frac{\alpha_s^2}{16\gamma^2(1\!-\!\gamma)^2}
              =\alpha_s^2 v(\gamma)v(1\!-\!\gamma)\ ,
\label{eqn:27}
\end{equation}
with
\begin{equation}
v(\gamma)=\frac{2^{-2\gamma-1}}{\gamma}\frac{\Gamma(1\!-\!\gamma)}{\Gamma(1\!+\!\gamma)}
\label{eqn:28}
\end{equation}
which is, up to a factor $\frac{\alpha_sN_c}{\pi}$, the eikonal coupling of a gluon to
a dipole \cite{nprw,these} at the lower vertex.
Thus, the factorized factor $\{...\}$ is nothing else than
$v(1\!-\!\gamma)$, i.e. the coupling of the gluon to the photon's dipole configuration,
as shown in figure 1.
\item Using (\ref{eqn:27},\ref{eqn:28}), equation (\ref{eqn:phi}) can be rewritten
\begin{equation}
A_{el}(\gamma)\cdot\phi_{T,L}(\gamma;M^2)=\alpha_{em}\frac{\alpha_{s}N_c}{4\pi}
v(\gamma)\cdot \frac{h_{T,L}(\gamma;M^2)}{\gamma}\ .
\label{eqn:relation}
\end{equation}
In the hard perturbative domain where we consider a photon-dipole interaction
with a dipole of small size, formula (\ref{eqn:relation}) means that both high energy factorization
and the wave-function formalism are identical. 
The cross section can be
equivalently factorized in two ways: either by the convolution of the photon gluon
cross section times the gluon coupling to the dipole (right hand side), 
or by the probability
distribution of a pair of quarks in the photon times the dipole-dipole elementary
interaction (left hand side).

However, as previously discussed, the non perturbative input
may lead to distinguishable models. 

\item At the level of the non perturbative input, we note a relation between
the two formulations (\ref{eqn:tot})
and (\ref{eqn:ftl}), namely
\begin{equation}
\omega(\gamma)=\frac{2\alpha_sN_c}{\pi}n_{eff}(\gamma)
  \frac{v(\gamma)}{\gamma}\ ,
\end{equation}
which generalizes the result for $M=0$, obtained in ref. \cite{nprw}.
\end{itemize}


\section{Phenomenology}

Following our theoretical discussions of the previous sections, we will consider
two definite models relying on different formulations of the residue function
$\omega(\gamma)$. On the one hand, the model 1, with 
\begin{equation}
\omega(\gamma)= C_1\;\;\mbox{(constant)}
\end{equation}
corresponds to the factorization at the gluon level
(see figure 1). 
On the other hand,
in the wave function formulation of the dipole model of section 4, we can
reformulate the integrand of the structure function 
(see equation (\ref{eqn:tot})) by using relation (\ref{eqn:relation}), namely:
\begin{equation}
\frac{h_{T,L}(\gamma)}{\gamma}\cdot\omega(\gamma)=\left(\frac{4\pi\alpha_s}{\alpha_{em}N_c}\right)
\phi(\gamma)\cdot \omega(\gamma)\cdot v(1\!-\!\gamma)\ .
\end{equation}
In the model 2, the hard vertex is thus described by factorizing 
$\phi(\gamma)$ which means that we consider
\begin{equation}
\omega(\gamma)\cdot v(1\!-\!\gamma)=C_2 \mbox{ (constant)}\ .
\label{eqn:ansatz}
\end{equation}
As expected, $\omega(\gamma)$ behaves like $\gamma$ when $\gamma$ goes to
zero in this framework.
In the large $|\gamma|$ region ($\gamma={1}/{2}+i\nu$, $\nu$ goes to infinity), 
one has $|\omega(\gamma)|\simeq|{C_2}/{v(1\!-\!\gamma)}|
\propto |\gamma|^2$. The exponential 
cutoff of $\phi_{T,L}$ is the same as the one (\ref{eqn:behav}) 
of the coefficient function and thus the large $|\gamma|$ 
constraint is satisfied.
Note that, in both cases, 
the proton structure functions depend on three free parameters
only: the global normalization $C_{1}$ or $C_2$,  
the effective constant strong coupling $\alpha_s$, and 
the non perturbative scale $Q_0$.

We determine these parameters for the two models 
by a fit of $F_2=F_T+F_L$ in their kinematical 
region of validity ($x\leq 10^{-2}$).
In this region, $Q^2$ is automatically limited by the HERA kinematics.
Using the corresponding 103 experimental points given
by the H1 collaboration \cite{f2}, we fit
our results with the contribution of the three light
quarks $u,d,s$ (assumed massless) 
and of the charm quark (mass $M_c$). 
As usual, we will vary
$M_c$ in the range $1.35-1.7\;\mbox{GeV}$
\cite{f2c}. 

The $F_2$-fit for the medium mass $M_c=1.5\;\mbox{GeV}$ 
is displayed in figure 2, together with the predictions for its charm
component $F_2^c$.
In table I, we give the $\chi^2$
and the value of the fit parameters for the fits for model 1 and 2 and 
for $M_c=1.35,\ 1.5,\ \mbox{and}\ 1.7$.
For model 1, the $\chi^2$ per point is always less than $0.9$, while for model 2
it is even lower.

Some comments on the parameter values are in order.
For fit 1, the value of $Q_0$ is around 330 MeV which is a typical non perturbative
scale for the proton. The value of the effective coupling constant in
the BFKL mechanism $\alpha_s$  ($0.07$) is rather low. This value would amount to an
effective pomeron intercept $\alpha_P=1+\alpha_s N_c 4\log 2/\pi \simeq 1.18$.
(It is known that $\alpha_P$
is influenced by sub-leading corrections to the BFKL kernel, 
see e.g. references \cite{cia}).
Note that the fit of model 1 for $F_2$ in the framework of 
the QCD dipole model and high energy factorization
is compatible with the previous ones in the same framework \cite{nprw}.

We also in parallel performed the
phenomenological analysis 
using the model 2-ansatz (\ref{eqn:ansatz}).
The obtained fit reproduces fairly
well the data for $F_2$ (see figure 2). 
As indicated by the lower value of $\chi^2$, it seems that the data,
especially in the small-$x$ small-$Q^2$ region are even better reproduced
than for model 1.
Note that the value of $Q_0\simeq 1.2$ is substantially higher
and the effective coupling constant $\alpha_s$ is larger 
($\simeq \alpha_s(M_Z)$).

Following the factorization scheme and
using formulae (\ref{eqn:finale}) (model 1) or (\ref{eqn:ftl}) (model 2),
we obtain now parameter-free predictions for the longitudinal structure functions
$F_L$ and $F_L^c$ (see figure 3).
The predictions for $F_L$ and $F_L^c$ are shown together with the
indirect H1 determination of $F_L$ \cite{fl}.
As expected, $F_L^c$ becomes a significant part of $F_L$ at small $x$ and
high $Q^2$. As already noticed \cite{nprw}, the predictions are low but compatible 
with the present large error bars. Note also that the determination \cite{fl}
depends on the theoretical scheme one considers \cite{thorne2}, so it is
difficult to draw any conclusion on $F_L$ at this stage.

The main outcome of our analysis is a parameter free prediction for $F_2^c$, the
charm component of the structure function (see fig.4).
When compared to ZEUS and H1 data for the charm component in
the same range in $x$ and $Q^2$ \cite{f2c}, 
we find a good agreement within the present experimental error bars.
The extrapolation of the prediction to the kinematical range of EMC data is
correct while the dipole model is not expected to be valid 
in this region. 

Looking in more detail to the predictions of model 1 (figure 4-a)
and model 2 (figure 4-b), we observe the following features.
The dispersion of the results with respect to $M_c$ 
is rather small for fit 1 with a maximum of 10\% when $x$ reaches $10^{-4}$. 
Moreover,
the prediction is comparable to the next-leading order GRV prediction
which proves that $F_2^c$ cannot allow one to distinguish between
the two approaches.

For model 2,
the predictions obtained for $F_2^c$ are displayed in figure 4-b. 
The prediction for the HERA region is
quite satisfactory and, interestingly enough, 
somewhat higher than both the NLO GRV predictions and fit 1.
It would certainly be useful to deserve some experimental and theoretical
attention to this difference. Indeed, the scheme of model 2 which
is based on a different factorization than both GRV and model 1 could
lead to a better understanding of the data.
A sensible decrease of the experimental errors on $F_2^c$
in the region where $Q^2$ is
moderate (of the order $10\ \mbox{GeV}^2$) and $x$ small
(of the order $10^{-3}$) could allow a refinement of this analysis.

All in all, the comparison of fits 1 and 2 shows that the high energy factorization
prediction for $F_2^c$ depends on the non-perturbative input in the
HERA range. Indeed, a different factorization scheme such as model 2
corresponds to a modification of the non perturbative input of high energy factorization.
However, the largest uncertainty due to this effect is of the order 20\%
and less than the experimental uncertainty. Thus, there is no
present evidence of a distinction between these schemes and the DGLAP scheme.


\section{Conclusions and outlook.}

Let us summarize the results of our analysis.\\
{\bf (i)} High energy factorization \cite{catani} gives predictions for the
$x$ and $Q^2$ behaviour of $F_2^c$ at low $x$ in the framework of 
models based on the BFKL dynamics. 
The charm prediction is fixed by high energy factorization once
the total proton structure function $F_2$ is fit. More generally,
this prediction is well defined for any quark mass.\\
{\bf (ii)} The factorization of the hard virtual photon vertex at the level
of the exchanged gluon (model 1) or the exchanged quark (model 2) 
leads to two specific classes 
of models which both give a satisfactory fit of $F_2$.
The formulation \cite{bjorken,nikolaev} 
based on the virtual photon wave function is shown to lead to a
physical realization of model 2.\\
{\bf (iii)} Both models lead to predictions for $F_2^c$ in agreement with the
present data. However, the second scheme leads to a higher $F_2^c$
than both model 1 and NLO DGLAP predictions \cite{f2c}. 
This justifies further
experimental and theoretical studies.\\
{\bf (iv)} The derivation leads to a new constraint on the unintegrated
gluon distribution in the proton 
due to the renormalization group evolution at
high transverse momentum of the off-shell gluon.\\

It is interesting to discuss the gluon distribution functions
in this framework. In the high energy factorization scheme, it would consist in
an expression like (\ref{eqn:tot}), replacing the coefficient functions by unity.
However, the convergence of the integral (\ref{eqn:tot}) is then not {\it a priori} preserved and
even if convergent, the dependence of the result on the large $|\gamma|$ region
would be larger, casting a doubt on the relevance of the BFKL dynamics.
In the wave-function framework, the extraction of the gluon structure function
is even more problematic since the factorization is at the quark level (see fig.1).
On a more general ground, this confirms the statement of caution \cite{catanew}
about extracting the gluon from this kind of analysis,
especially at low $x$.
We think that this point deserves more studies.

Finally, the running of the coupling constant and other aspects
of next leading order BFKL resummation features have been neglected 
in the present analysis. The small value obtained for the effective coupling
constant in the fits clearly indicates that such effects are important and should
be included in a more detailed analysis. Indeed, a preliminary theoretical hint
on the effective behaviour of the pomeron singularity at NLO accuracy leads
to a small and constant intercept of the order $0.2$ \cite{cia}.
A NLO BFKL analysis of $F_2^c$ would thus
be required in view of a future improvement of the experimental analysis.\\

\noindent
{\large\bf Acknowlegments}

We thank Stefano Catani for fruitful discussions and Henri Navelet and Christophe Royon
for stimulating remarks.

\eject

\newpage

\noindent
{\bf TABLE CAPTION}

\vspace{1cm}
\noindent
{\bf Table I}

\vspace{0.5cm}
{\it Total proton structure function: 
$\chi^2$ and parameters for the fits}\\

For each model, we give the $\chi^2$ value for 103 points
and the parameters $C$, $Q_0$ and $\alpha_s$. Three different
charm masses, $M_c=1.35,1.5,1.7\ \mbox{GeV}$
are considered. {\bf I-a.} Model 1. {\bf I-b.} Model 2.

\clearpage
\begin{center}
\epsfig{file=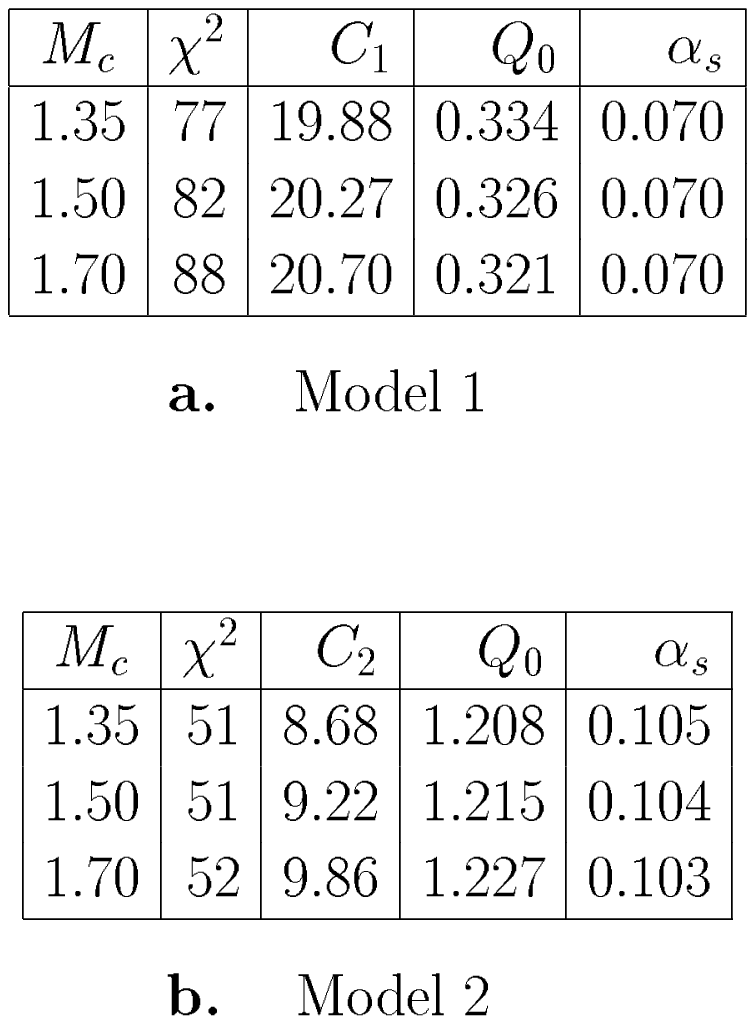}
\end{center}\begin{center}
{\large\bf Table I}
\end{center}
\clearpage

\noindent
{\bf FIGURE CAPTIONS}
 
\vspace{1cm}
\noindent
{\bf Figure 1} {\it Quark-antiquark pair leptoproduction at high energy}\\

The upper vertex represents the virtual photon (with momentum $q$)-virtual gluon 
(with momentum $k$) fusion diagram through the production
of a pair of quarks of mass $M$.
The proton (with momentum $p$) interacts with the gluon through the BFKL kernel.
The non perturbative proton vertex is schematized by the shaded area.
Model 1 corresponds to factorization at the gluon level, model 2 at the quark level,
see text.

\vspace{0.5cm}
\noindent
{\bf Figure 2}
{\it The fits (model 1 and model 2) for the structure function $F_2$}\\

The structure function $F_2$ and its parametrization are displayed for model 1 and model 2.
The fits have been performed with the 1994 H1 data (triangles)
but only using the 103 experimental points for which $x\leq 10^{-2}$. 
Continuous line: fit for model 1;
dashed line: fit for model 2;
dotted line: prediction for the charm component
$F_2^c$, for $M_c=1.5\;\mbox{GeV}$ and model 1. 
The available H1 data are marked by stars.

\vspace{0.5cm}
\noindent
{\bf Figure 3}
{\it Predictions for the longitudinal structure functions $F_L$ (continuous line) 
and $F_L^c$ (dashed line)}\\

Continuous line: model 1; dashed line: model 2; dotted line: $F_L^c$ (model 1).
The experimental points available from H1 \cite{fl} have been reported 
on the graph.

\vspace{0.5cm}
\noindent
{\bf Figure 4}
{\it Predictions for $F_2^c$}\\

The reported data on the plots are from H1 $D^0$ (squares), H1 $D^*$ (circles),
ZEUS $D^*$ (full circles), and at lower energy, EMC data (crosses).
The predictions of our models are displayed by a band delimited
by the two continuous lines ($M_c=1.35$ for the higher curves and 
$M_c=1.7$ for the lower curves).
Extrapolations of our predictions down to the EMC range are plotted as
dot-dashed lines.\\
Figure {\bf 4-a}: model 1 predictions.
The GRV prediction \cite{GRV} based on NLO analytic calculation
is indicated by the shaded band (borrowed from ref. \cite{f2c}, second paper).\\
Figure {\bf 4-b}: model 2 predictions: the band of solutions is delimited by thick lines.
The model 1 predictions are recalled by the band delimited by thin lines, 
for comparison.

\clearpage
\begin{center}
\epsfig{file=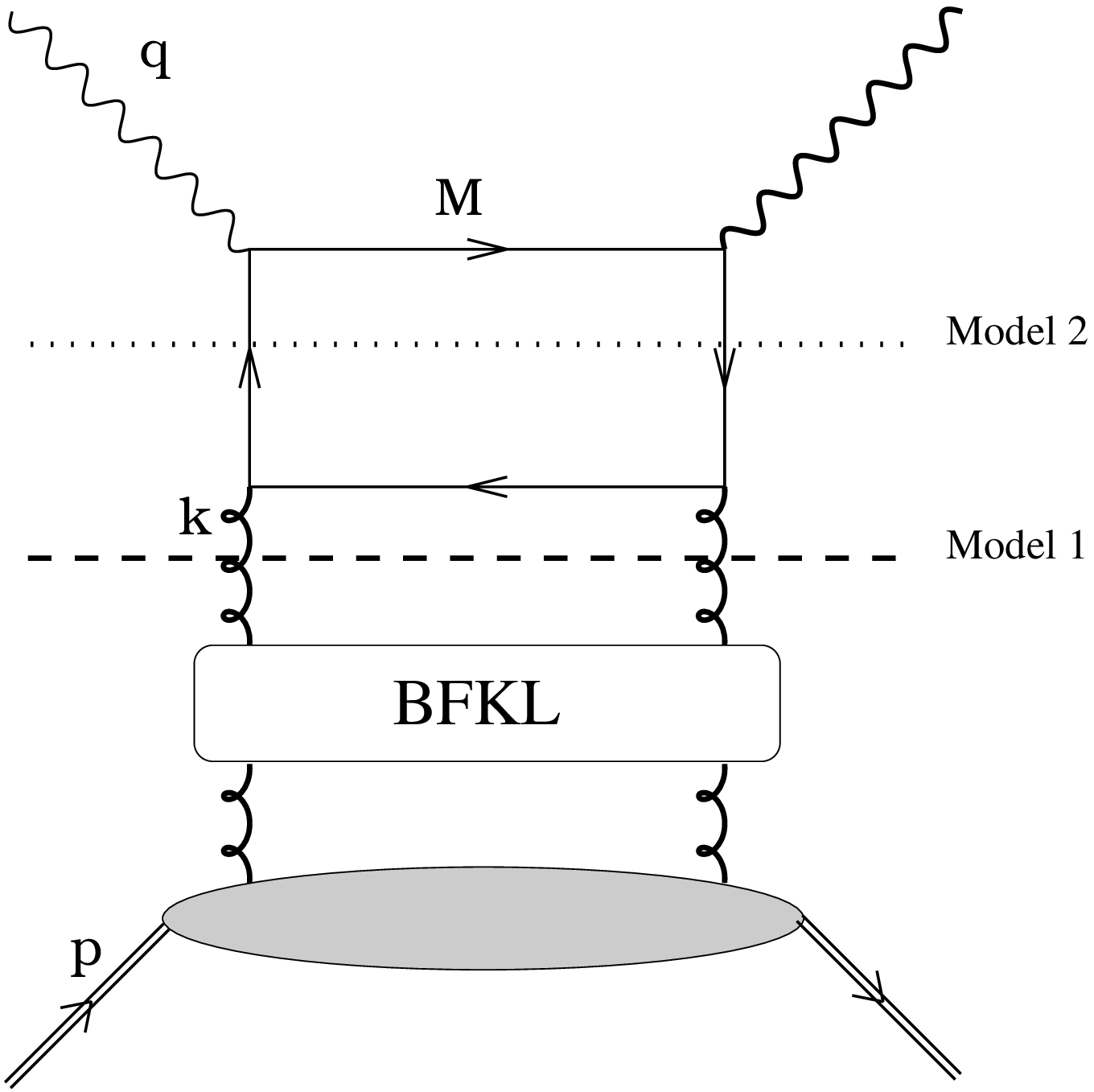}
\end{center}\begin{center}
{\large\bf Figure 1}
\end{center}

\clearpage
\begin{center}
\epsfig{file=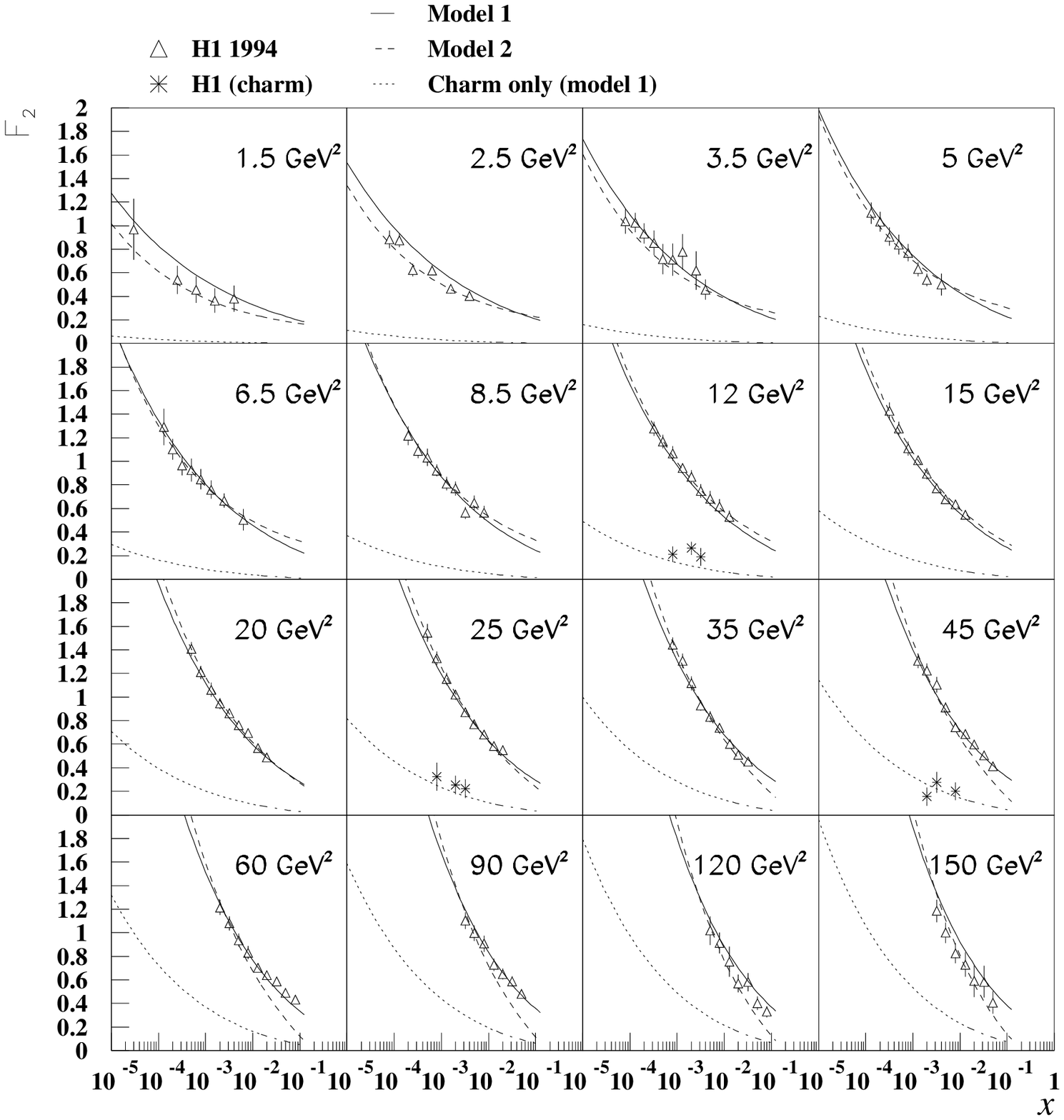,height=15cm}
{\large\bf Figure 2}
\end{center}

\clearpage
\begin{center}
\epsfig{file=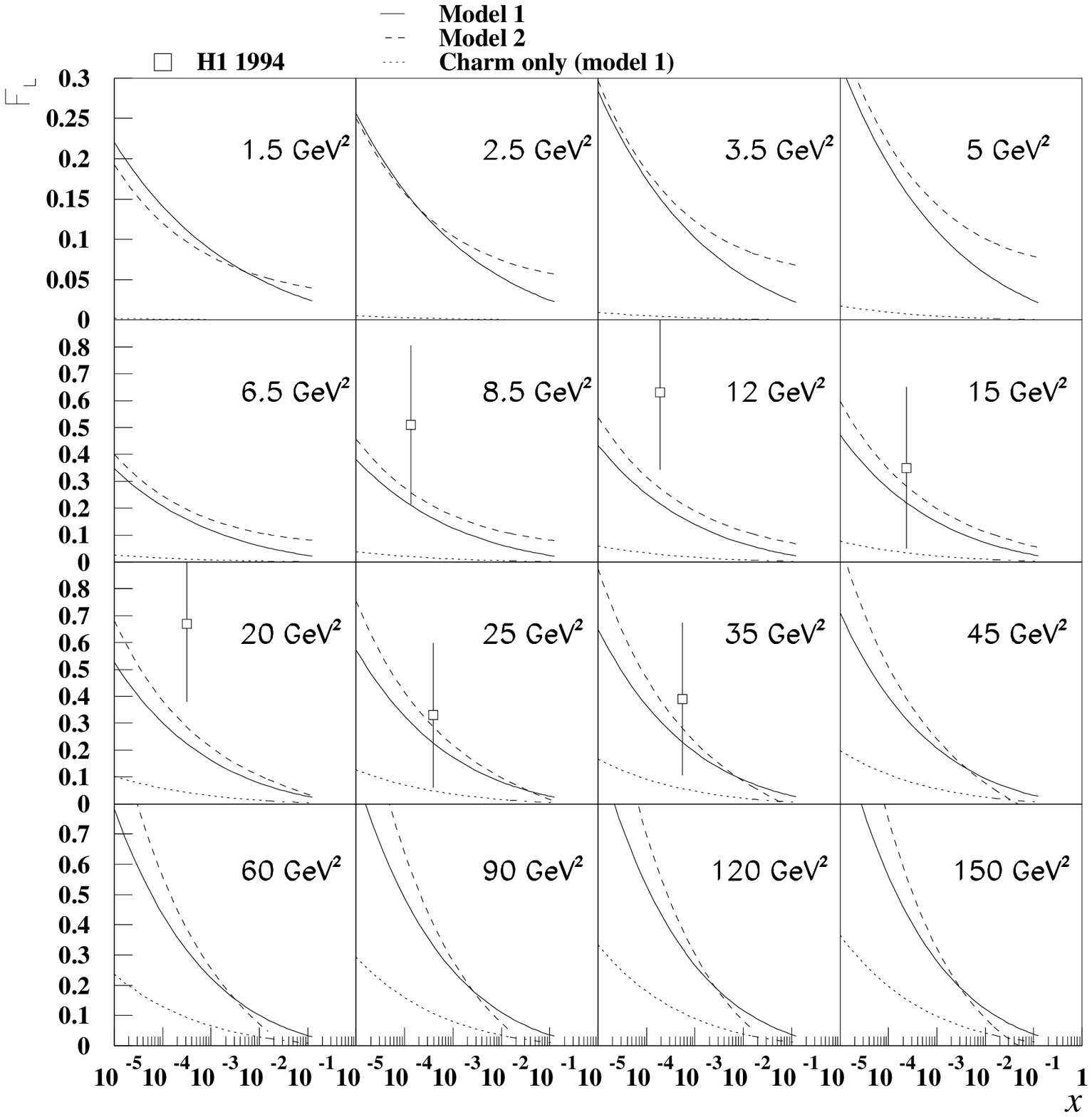,height=15cm}
{\large\bf Figure 3}
\end{center}

\clearpage
\begin{center}
\epsfig{file=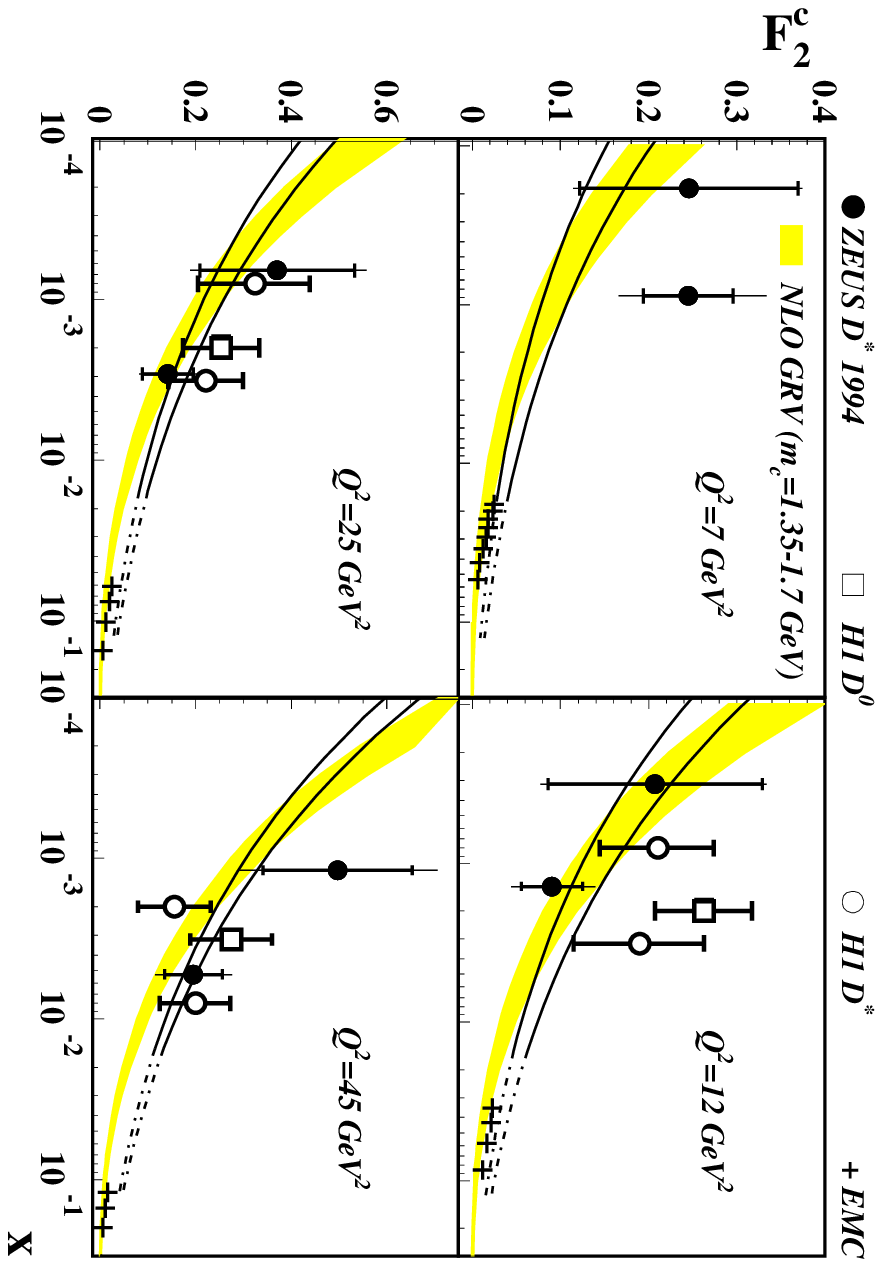,height=15cm}
\end{center}
\begin{center}
{\large\bf Figure 4-a}
\end{center}

\clearpage
\begin{center}
\epsfig{file=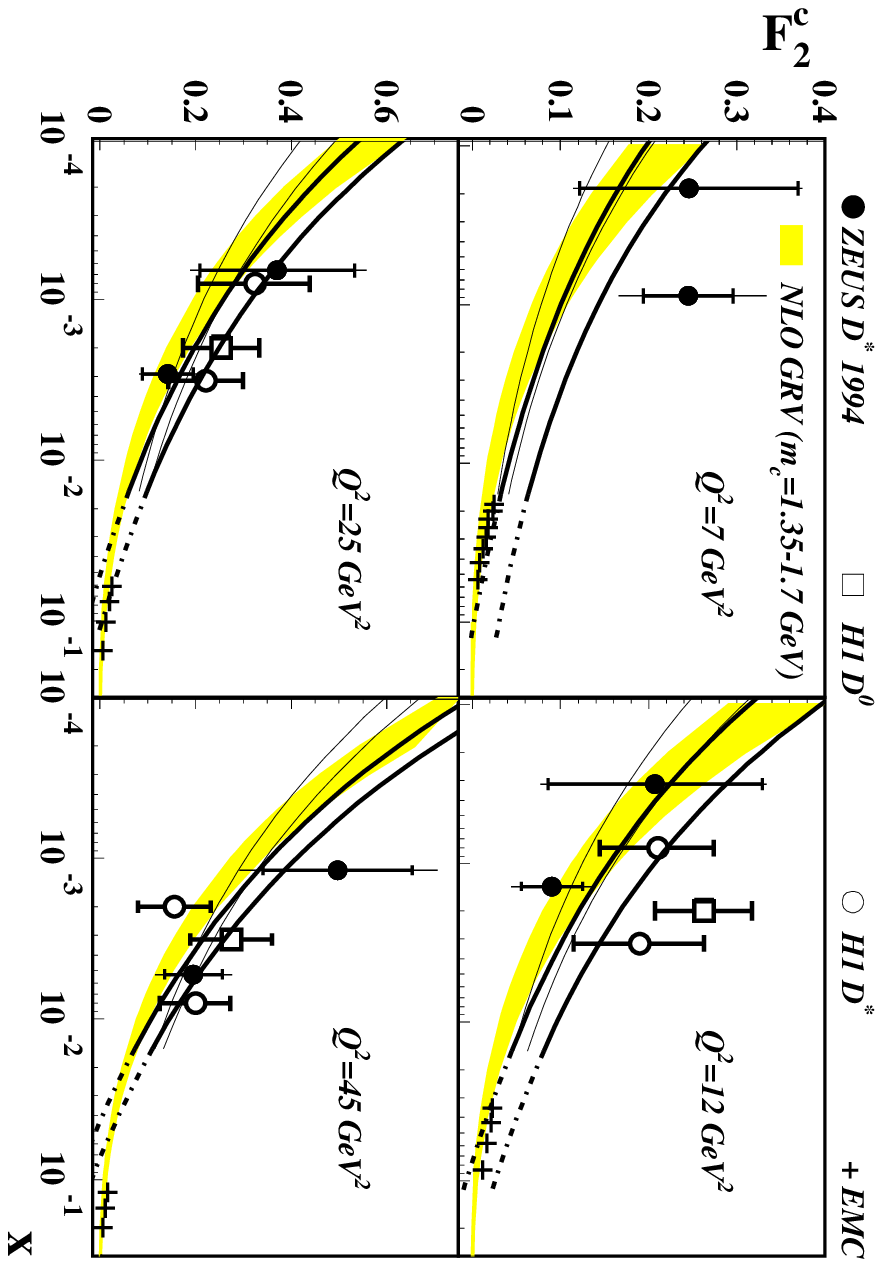,height=15cm}
\end{center}\begin{center}
{\large\bf Figure 4-b}
\end{center}

\end{document}